\documentclass[12pt,preprint]{aastex}
\usepackage{amsmath, amsthm, amssymb}
\usepackage{natbib}
\usepackage{lscape}

\ifnum 2<1
\newcommand{\aap}{{Astron. Astrophys.}}
\newcommand{\apj}{{Astrophys. J.}}
\newcommand{\apjl}{{Astrophys. J. Lett.}}
\newcommand{\araa}{{Annual Review of Astron and Astrophys}}
\newcommand{\grl}{{Geophys. Res. Lett.}}

\newcommand{\nat}{{Nature}}

\newcommand{\solphys}{{Solar Phys.}}

\newcommand{\ssr}{{Space Sci. Rev.}}
\fi

\begin{document}

\title{The cause of the weak solar cycle 24}

\author{J. Jiang\altaffilmark{1}, R.H. Cameron\altaffilmark{2},
        M. Sch\"{u}ssler\altaffilmark{2}}
\altaffiltext{1}{Key Laboratory of Solar Activity, National
Astronomical Observatories, Chinese Academy of Sciences, Beijing
100012, China}
\altaffiltext{2}{Max-Planck-Institut f\"ur Sonnensystemforschung,
               Justus-von-Liebig-Weg 3, 37077 G\"ottingen, Germany}

\email{jiejiang@nao.cas.cn}

\begin{abstract}
{The ongoing 11-year cycle of solar activity is considerably less
vigorous than the three cycles before. It was preceded by a very
deep activity minimum with a low polar magnetic flux, the source of
the toroidal field responsible for solar magnetic activity in the
subsequent cycle. Simulation of the evolution of the solar surface field
shows that the weak polar fields and thus the weakness of the present
cycle 24 are mainly caused by a number of bigger bipolar regions emerging at
low latitudes with a
`wrong' (i.e., opposite to the majority for this cycle) orientation of
their magnetic polarities in the North-South direction, which impaired
the growth of the polar field. These regions had a particularly strong
effect since they emerged within $\pm10^\circ$ latitude from the solar
equator.}
\end{abstract}


\newpage
\section{Introduction}

Solar activity, as manifested by sunspots and bipolar magnetic regions
at the solar surface, by energy release in flares, eruptions, coronal
mass ejections, and energetic particles, varies cyclically with a
(quasi-)period of about 11 years \citep{Hathaway2010}. Activity cycles
differ significantly with regard to their length and strength. The
present cycle is among the weaker ones,
particularly if compared with the sequence of strong cycles during the
second half of the 20th century. Cycle 24 was preceded by an unusually
extended period of very low activity \citep{Russell2010}, weak polar
magnetic field \citep{Munoz2012}, and low heliospheric open flux
\citep{Smith2008, Owens2013}. This is consistent with the empirical
correlation of the polar field around solar activity minimum with the
strength of the subsequent cycle \citep{Wang2009}. This correlation
reflects the role of poloidal magnetic flux connected to the polar
fields as the dominant poloidal source of the toroidal flux generated by
differential rotation in the solar interior \citep{Cameron2015}, whose
emergence at the surface during the subsequent 11-year cycle leads to
the various manifestations of solar activity.

Generally, the observed large-scale solar surface field is well
represented by Surface Flux Transport (SFT)
simulations \citep{Wang1989,Sheeley2005,Mackay2012,Jiang2014a}, which
address the evolution of the magnetic field as a result of the emergence
of bipolar magnetic regions (sunspot groups) and the subsequent
transport of magnetic flux by near-surface plasma flows (convection,
differential rotation, poleward meridional flow). Such models showed
that the polar fields are reversed and built up by preferred flux
transport of one polarity across the equator as a result of the systematic
tilt of bipolar sunspot groups with respect to the direction of
rotation. The strength of the polar fields mainly depends on the tilt
angle, the distribution of emerging flux in heliographic latitude, and
the speed of the meridional flow \citep{Baumann2004}. SFT simulations
reproduce the polar fields (as regularly measured directly since 1976
and inferred from geomagnetic and solar proxies measured since the 19th
century) relatively well for the cycles before cycle 23.  However,
unless they are continuously adjusted by data assimilation from
observations \citep{Upton2014a}, such simulations so far could not
reproduce the very weak polar field during the declining phase and
activity minimum of cycle
23 \citep{Jiang2010,Yeates2014,Upton2014b}. Suitable ad-hoc variations of
the meridional flow were considered as a possible
remedy \citep{Schrijver2008,Wang2009,Nandy2011,Jiang2013}, but there is no
empirical basis for changes of the kind required by these
models \citep{Gonzalez2008,Hathaway2010b,Cameron2010a}.

While systematic variations of the tilt angles of sunspot groups
\citep{Cameron2010b} were found to be an unlikely cause of the weak
polar fields of cycle 23 \citep{Schrijver2008,Jiang2013}, there is
evidence for a considerable random element in the evolution of the polar
fields: the observed scatter of the tilt angles causes a variation of
30--40\% (standard deviation from the mean for many realizations) in the
resulting polar field around activity minima
\citep{Baumann2004,Jiang2014b}.  Even single big bipolar regions
emerging very near or across the solar equator can significantly affect
the amplitude of the polar field \citep{Cameron2013}. This suggests that
an answer to the question why the polar fields in the declining phase of
cycle 23 were so weak (and, as a consequence, why cycle 24 is rather
feeble) requires a study that includes detailed information about the
individual tilt angles and magnetic polarities of the bipolar regions
that emerged during cycle 23.  Such data became available only recently
\citep{Li2012,Stenflo2012}, so that, in contrast to previous studies, we
could run SFT simulations including the actual tilt angles for cycle 23.

\section{Surface flux transport simulations}

For our simulations we used the SFT code described in
\citet{Baumann2004} and \citet{Cameron2010b}, which solves the
magnetohydrodynamic induction equation describing the evolution of the
radial component of the large-scale magnetic field at the solar surface,
$B(\lambda,\phi,t)$ (where $\lambda$ and $\phi$ are heliographic
latitude and longitude, respectively), as a result of passive magnetic
flux transport by surface flows, viz.
\begin{eqnarray}
\label{eqn:SFT} \nonumber\frac{\partial B}{\partial t}=& &
-\Omega(\lambda)
                       \frac{\partial B}{\partial \phi}
         - \frac{1}{R_\odot \cos\lambda}
              \frac{\partial}{\partial \lambda}[\upsilon(\lambda)
         B \cos \lambda] \\ \noalign{\vskip 2mm}
& & +\eta \left[\frac{1}{R_\odot^2 \cos{\lambda}}
                \frac{\partial}{\partial \lambda}\left(\cos\lambda
          \frac{\partial B}{\partial \lambda}\right) +
     \frac{1}{R_\odot^2 \cos^2{\lambda}}\frac{\partial^2 B}{\partial
     \phi^2}\right] + S(\lambda,\phi,t).
\end{eqnarray}
The latitudinal differential rotation of Sun, $\Omega(\lambda)$, is
taken from \citet{Snodgrass83}.
For the poleward meridional flow, $\upsilon(\lambda)$, we used the profile by \citet{Ballegooijen98},
which is consistent with the empirical profile derived by magnetic
element tracking \citep{Hathaway2011}. The
``turbulent'' magnetic diffusivity of $\eta=250$~km$^2$s$^{-1}$
describes the random walk of the magnetic flux elements in the
large-scale convective flow pattern of supergranulation
\citep{Leighton1964}.

The source term, $S(\lambda,\phi,t)$, represents
the emergence of bipolar magnetic regions. As input database for these
we used the sunspot observations obtained by the Solar Optical Observing
Network (SOON) of the US Air Force, which yields 3046 bipolar regions
between June 1996 and December 2010 (2921 regions during cycle 23 until
the end of 2008). The corresponding magnetic flux is determined by a single
parameter, $B_{\textrm{max}}$, which is calibrated by the total unsigned
surface flux obtained from SoHO/MDI synoptic maps after
rebinning to the spatial resolution of the simulation (1$^\circ$ in
both latitude and longitude) \citep{Scherrer1995}.
The match of the observation and simulation during the activity
maximum of cycle 23 yields $B_{\textrm{max}}$ = 592G. We applied the procedure described in
\citet{Baumann2004} to define the properties of the corresponding
bipolar regions. Further details of the source generation and
calibration are given in \citet{Cameron2010b}.

For the tilt angles of the emerging bipolar regions we used the data set
of  \citet{Li2012}, which is based upon daily magnetograms from the Mount
Wilson Observatory and from the SoHO/MDI. These data provide superior information about tilt angles since
they contain magnetic polarity information and also include plage
regions, both of which are not available when considering only
white-light images of sunspot groups \citep{Wang2015}. As in our
previous studies, the tilt angles were multiplied by a factor 0.7 in
order to account for the near-surface inflows towards active regions
\citep{Gizon2010}.

Since single big active regions can have a significant impact on the
polar fields \citep{Cameron2013}, the input data were double-checked,
thereby eliminating 22 duplicate (recurrent) active regions, which would
otherwise have entered the simulation more than once -- a list of these regions is provided in Table \ref{tab:repeatingARs}.

As a comparison case, we also carried a SFT simulation that assigns to
each emerging bipolar region a tilt angle, $\alpha$, according to a
function of latitude (commonly referred to as Joy's law) of the form
$\alpha = T\sqrt{|\lambda|}$, where the constant $T =\sum_i
\alpha_i/\sum_i \sqrt{|\lambda_i|}=1.4$ and summation is over all
bipolar regions of cycle 23 \citep{Cameron2010b}.

Our SFT simulations start at the end of June 1996, around the minimum
of cycle 22, using as initial condition the (interpolated and
polar-field corrected) synoptic map of the surface magnetic field for
Carrington rotation 1911 obtained by SoHO/MDI \citep{Sun2011}. The
evolution of the large-scale field resulting from flux emergence in
bipolar magnetic regions and flux transport by surface flows according
to Eq.~(1) is then followed until the end of 2010.

\begin{figure}
\begin{center}
\includegraphics[scale=0.8]{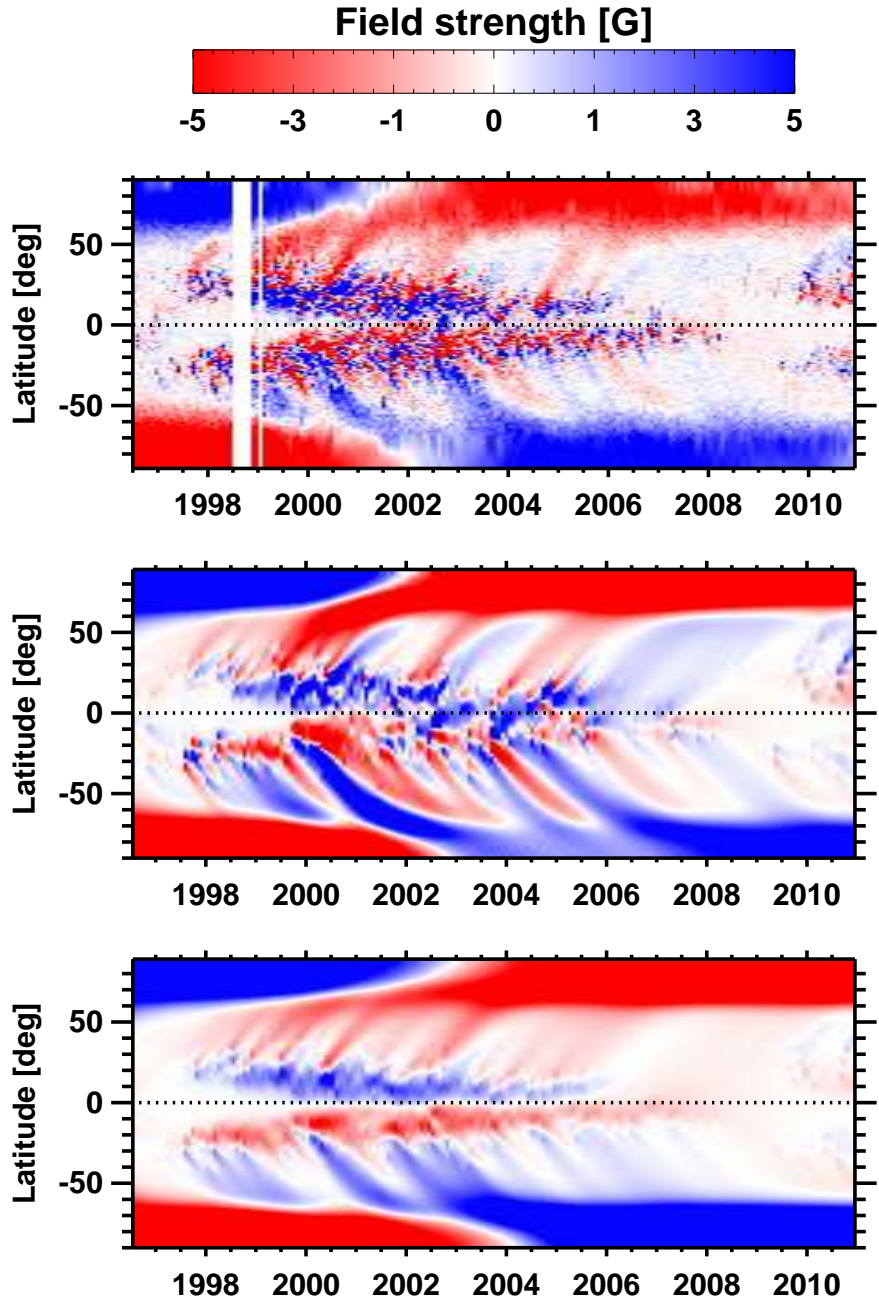}
\caption{Time-latitude diagrams of the longitudinally averaged radial
         magnetic field at the solar surface. Top panel: observation
         (SoHO/MDI synoptic magnetic field data); middle panel:
         simulation using the actual tilt angles of bipolar magnetic
         regions; bottom panel: simulation using tilt angles according
         to a fitted latitude dependence.}
\label{fig:butterfly}
\end{center}
\end{figure}

Fig.~\ref{fig:butterfly} shows observed and simulated time-latitude
diagrams of the longitudinally averaged radial surface field. Although
there are differences in detail, the simulation on the basis of the
actual individual tilt angles (middle panel) reproduce the main features
of the evolution of the surface field rather well. Deviations in detail
can be traced back to a small number of errors in the tilt data of
\citet{Li2012}. We have confirmed that these errors do not significantly
affect the polar fields and dipole moment around the activity minimum,
for which only the cross-equator transport of magnetic flux from
low-latitude active regions is important \citep{Jiang2014b}. To avoid
potential bias, we therefore chose not to correct these errors, which
would have improved the agreement of the time-latitude diagrams.

The main difference between the simulation with actual tilt angles and
that with Joy's law becomes apparent when considering the axial dipole
moment of the Sun,
\begin{equation}
D(t)=\frac{3}{2} \int_{\pi/2}^{-\pi/2} \left\langle B(\lambda,t)\right\rangle
              \cos\lambda \sin\lambda \, d\lambda,
\end{equation}
where $\langle B(\lambda,t)\rangle$ is the longitudinally averaged
radial surface field.  The axial dipole moment is a better suited
quantity than the polar field for the comparison of observation and
simulation. It is uniquely defined and limits the effect of asymmetries
in the magnetic flux distribution on the two hemispheres (such as
non-synchronous reversals of the polar fields). Around
activity minima, the axial dipole moment is dominated by the polar
fields and represents the relevant quantity for the poloidal magnetic
flux connected to the polar fields. Likewise, it also represents the
open heliospheric magnetic flux during these periods.

\begin{figure}
\begin{center}
\includegraphics[scale=0.9]{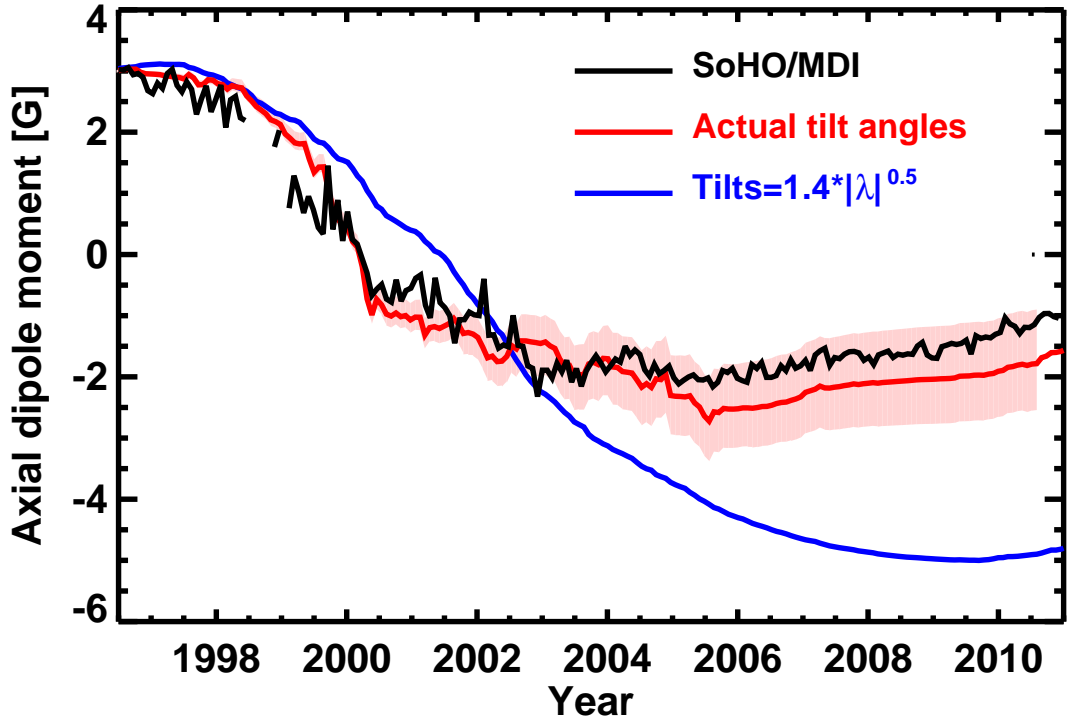}
\caption{Time evolution of the solar axial dipole moment. The
         curves correspond, respectively, to observed SoHO/MDI magnetic
         maps (black), a simulation using the actual tilt angles of
         bipolar magnetic regions (red), and a simulation using tilt
         angles according to a fitted latitude dependence (blue).}
\label{fig:dipole}
\end{center}
\end{figure}

Fig.~\ref{fig:dipole} shows the evolution of the axial dipole moment as
determined from observations (SoHO/MDI magnetic maps) and our SFT
simulations. The simulation based on the actual tilt angles follows the
observed evolution of the axial dipole moment well within the shaded
error range (see below).  On the other hand, employing Joy's law for the
tilt angles leads to a much too strong axial dipole moment in the
declining and minimum phases of cycle 23. This behavior is typical for
previous attempts to reproduce the observed weak axial dipole moment and
polar field by SFT simulations, unless the models were arbitrarily
modified (e.g., in terms of meridional flow variations).

We estimate the error of the axial dipole moment in our simulations as
follows. The main source of uncertainty are the errors in the
determination of the tilt angles. We have therefore used a different
data source, the Kitt Peak Synoptic Magnetograms, to independently
redetermine the tilt angles of about 30\% of the bipolar regions
considered by \citet{Li2012}, who used SoHO/MDI magnetograms. The
regions were chosen from the SOON list, their area determined by
applying a threshold on measured magnetic flux density, and their tilt
then calculated using the line connecting the centers of gravity of the
positive and negative polarity parts. Comparison with the data of
\citet{Li2012} yields an overall value of $\sigma=11.5^\circ\,$ for the
rms scatter. However, the error is dependent on the area of the region;
e.g., for the 50 biggest ARs, the scatter is reduced to about
$5^\circ$. The dependence on area, $A$ (in microhemispheres), can be
roughly approximated by the function
\begin{equation}
\sigma(A) =\left\{
  \begin{array}{l l}
     18^\circ - 0.02 A & \textrm{for} A<650. \\
     5^\circ & \textrm{otherwise}.
  \end{array}
  \right.
\end{equation}
We carried out 20 SFT simulations with random scatter of the AR tilts
according to this function. The rms variation of the resulting axial
dipole strength is then taken as error estimate.

\section{Origin of the low dipole moment}

Figure~\ref{fig:cumulative} illustrates the reason for the discrepancy
between the simulation with actual tilt angles and the one based upon
Joy's law.  Shown are cumulative contributions of magnetic regions to
the change of the dipole moment (since 1996) as a function of emergence
latitude (modulus). The individual contribution of a magnetic region for
large times is proportional to $\Delta =
A^{3/2}\,\sin(\alpha)\exp(-\lambda^2/110)$, where $A$ is the area of the
region \citep{Jiang2014b}. The normalized cumulated contributions using
the actual tilt angles (dashed curve) closely follow those when assuming
Joy's law (solid curve) for emergence latitudes poleward of
$\pm10^\circ$ from the equator. Nearer to the equator, the curves
diverge: the actual tilt angles of magnetic regions emerging in this
range statistically deviate strongly from Joy's law in such a way that
the resulting total change of the dipole moment since 1996 is reduced by
about 40\%. The is mainly due to comparatively few regions in the
(positive and negative) tails of the distribution of $\Delta$, i.e., a
result of random variations together with low-number statistics.

\begin{figure}
\begin{center}
\includegraphics[scale=0.9]{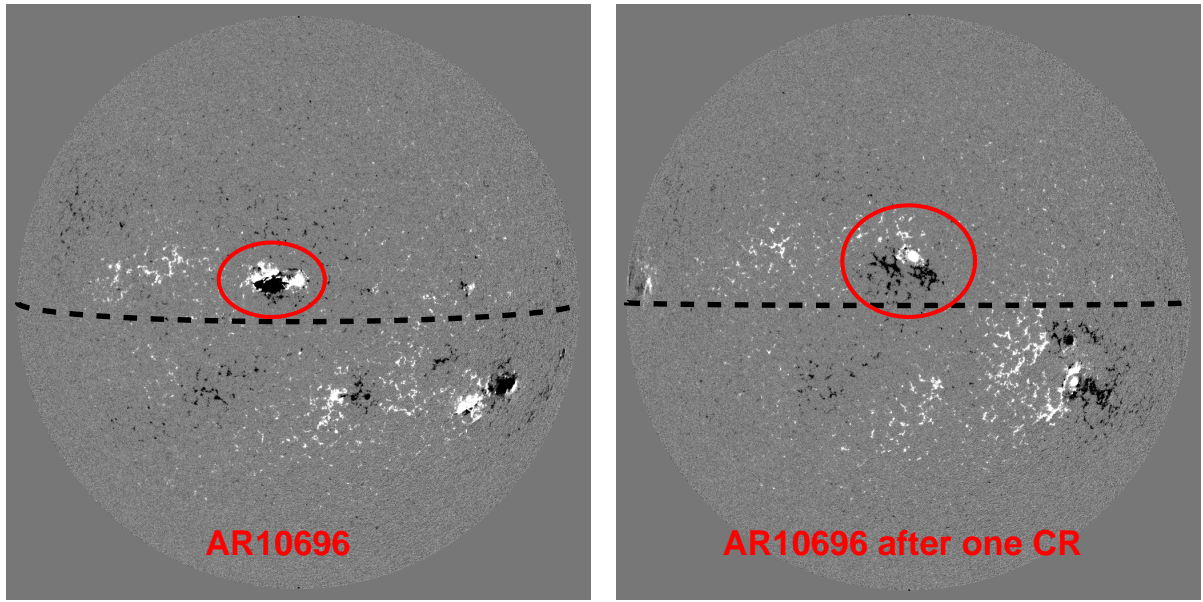}
\caption{Example of a bipolar magnetic region that significantly
  weakened the axial dipole moment in the declining phase of cycle 23.
  Shown are SoHO/MDI magnetic maps of the active region AR10696 taken
  Nov 5, 2004 (left panel) and Dec 2, 2004 (right panel, after one solar
  rotation, then denominated AR10708), respectively. Positive magnetic
  flux is indicated in white, negative flux in black. Owing to its
  near-equator emergence, high tilt, and abnormal polarity orientation
  in the North/South direction, the region provides a significant amount
  of negative flux that is transported over the equator (by
  supergranular random walk) to the southern hemisphere. Through
  poleward advection (mainly by meridional flow) this flux eventually
  weakened the buildup of positive flux around the south pole of the
  Sun, thus lowering the axial dipole moment.}
\label{fig:AR10696}
\end{center}
\end{figure}

An example for a big magnetic region with `wrong' orientation (i.e.,
opposite to the majority of the magnetic regions during this cycle) of
the magnetic polarities in the North-South direction is active region
AR10696, which appeared in November 2004. Magnetic maps for this region
are shown in Fig.~\ref{fig:AR10696}. Since it emerged near the equator,
a significant amount of negative-polarity magnetic flux could cross the
equator and thus significantly reduce the buildup of the axial dipole
moment in the descending phase of the cycle \citep{Cameron2013}. Since the
magnetic flux at the poles around activity minima is typically of the
order of the flux of only one big bipolar magnetic region, a few bipolar
regions emerging near the equator can have a significant impact.

\section{Comparison with cycles 21 and 22}

\begin{figure}
\begin{center}
\includegraphics[scale=0.9]{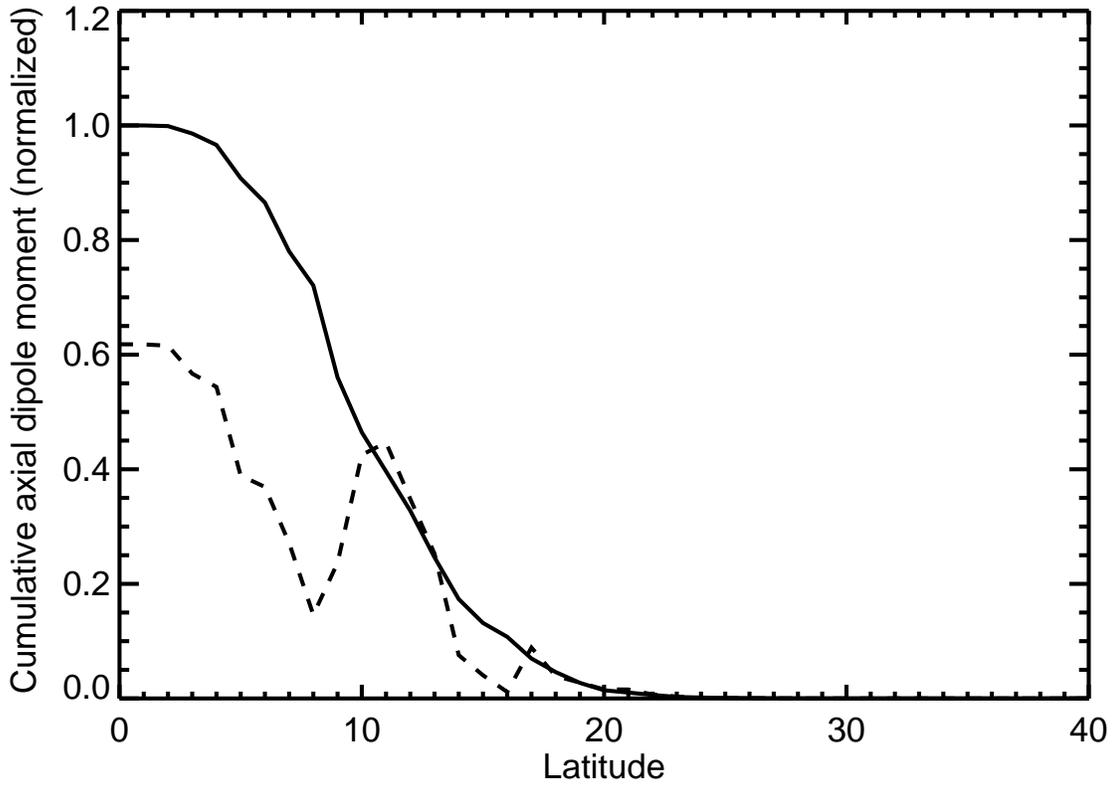}
\caption{Cumulative contributions (in latitudinal distance from the
         equator) of magnetic regions to the change of the axial dipole
         moment since 1996.  Results are given assuming tilt angles
         according to Joy's law ($\propto 1.4\sqrt{|\lambda|}$, solid
         curve) and for the actual individual tilt angles (dashed
         curve).}
\label{fig:cumulative}
\end{center}
\end{figure}

\begin{figure}[ht!]
\includegraphics[scale=0.6]{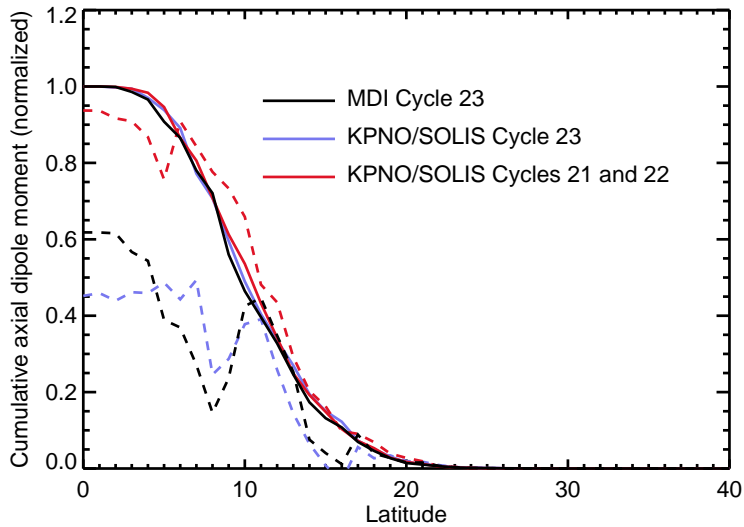}
\caption{Cumulative contributions (in latitudinal distance from the
         equator) of magnetic regions to the change of the axial dipole
         moment during cycles 21, 22, and 23. Full lines give the result
         assuming tilt angles according to Joy's law (almost coincident
         with each other), dashed lines for the case using the actual
         individual tilt angles. The black curves are reproduced from
         Fig.~\ref{fig:cumulative} (on the basis of MDI full-disk
         magnetograms), blue lines indicates results from KPNO synoptic
         magnetograms for cycle 23, and red lines the corresponding
         results for cycles 21 and 22 combined.}
\label{fig:cumulative_21_22_23}
\end{figure}

While Joy's law was a reasonable assumption in SFT simulations before
cycle 23, we have seen that it is inadequate to describe the evolution
of the axial dipole moment in cycle 23. It is therefore of interest to
consider a similar analysis as the one resulting in
Fig.~\ref{fig:cumulative} for earlier cycles. Since magnetic information
is necessary to properly determine the tilt angle (particularly for the
big regions with `wrong' tilt), we have to restrict ourselves to cycles
21 and 22, for which Kitt Peak National Observatory (KPNO) synoptic
magnetograms are available. Since these are based on observations along
the central meridian only, they contain not all of the emerging bipolar
regions detectable by full-disk magnetograms such as those from
SoHO/MDI, but still provide a statistically meaningful sample. We used
the synoptic magnetograms to determine the individual tilt angles of
bipolar regions emerging in cycles 21-23.  The (normalized) cumulative
distributions to the change of the dipole moment during these cycles are
given in Figure~\ref{fig:cumulative_21_22_23}. For comparison, the result
for cycle 23 on the basis of tilts from the MDI magnetograms
(Fig.~\ref{fig:cumulative}) is also reproduced.  The two results for
cycle 23 (from MDI and KPNO data, respectively) are consistent with each
other (taking into account that the KPNO analysis covers only 30\% of
the MDI regions): both show the big deviation from Joy's law for
latitudes below $10^\circ$. On the other hand, the result for cycle 21
and 22 (combined in order to improve statistics) is fully consistent
with tilt angles according to Joy's law.  This indicates these cycles
were less strongly affected than cycle 23 by randomness concerning the
tilt angles of low-latitude active regions in the descending phase of
the cycle.

\section{Concluding remarks}

Our SFT simulations have shown that the weak axial dipole moment (weak
polar fields) around the activity minimum of cycle 23 are a result of a
number of low-latitude bigger active regions whose `wrong' tilt angles
did not follow Joy's law. Since the poloidal magnetic flux related to
the axial dipole moment (directly related to the polar fields) is the
dominant source of the toroidal flux that emerges in the subsequent
cycle \citep{Cameron2015}, its low amplitude during the declining phase
of cycle 23 caused the weakness of the current cycle 24. The dependence
on the properties of big bipolar regions emerging in the vicinity of the
solar equator, which are affected by randomness, strongly limits the
scope of predicting solar cycle strength: until the amplitude of the
polar fields is established around solar minimum, no sensible prediction
of the strength of the next cycle can be made. Sound predictions
projecting even further into the future are virtually impossible.

\begin{acknowledgements}
The authors are grateful to Jing Li for providing the tilt angle data in
electronic form. SoHO is a project of international cooperation between
ESA and NASA. The SOON data were downloaded from
http://solarscience.msfc.nasa.gov/greenwch.shtml, the SoHO/MDI synoptic
maps from http://solarscience.msfc.nasa.gov/greenwch.shtml, and the KPNO
synoptic maps from http://diglib.nso.edu/ftp.html
This work utilizes SOLIS data
obtained by the NSO Integrated Synoptic Program (NISP), managed by the
National Solar Observatory, which is operated by the Association of
Universities for Research in Astronomy (AURA), Inc. under a cooperative
agreement with the National Science Foundation.

J. Jiang acknowledges support by National
Science Foundation of China (grants 11173033, 11221063, 2011CB811401)
and support by the Strategic Priority Research Program of the Chinese
Academy of Sciences (grant XDB09040200).
\end{acknowledgements}

\bibliographystyle{apj}

\newpage

\begin{table}[ht!]
\begin{center}
\caption{List of the recurrent ARs. At {\it Time (yyyymmdd)},
  the ARs reached their maximum area (in microhemispheres).  The column
  {\it AR} gives the active regions that are retained in the SFT
  simulations. }
\begin{tabular}{ccccccc}
\hline\hline
No. &  Recurrent AR  &  AR  & Time & Latitude  & Area   & Tilt \\
\hline
 1    & 8375  & 8398 &19981109  & 21.0  &  964  &48.13\\
 2    & 8771  & 8739 &19991125  &-15.0  & 1097  & 3.63\\
 3    & 8759  & 8731 &19991109  & 10.0  & 1230  & 1.54\\
 4    & 9085  & 9046 &20000720  & 15.0  &  405  &29.88\\
 5    & 9131  & 9090 &20000825  & 15.0  &  312  &38.84\\
 6    & 9636  & 9601 &20010926  & 14.0  &  538  &36.56\\
 7    & 9670  & 9628 &20011019  &-18.0  &  711  &16.66\\
 8    & 9672  & 9704 &20011026  &-18.0  &  791  &44.88\\
 9    & 9684  & 9715 &20011104  &  5.0  &  737  &16.04\\
10    & 9751  & 9715 &20011226  &  4.0  &  671  &15.86\\
11    & 9973  & 9934 &20020529  &-16.0  & 1283  & 8.43\\
12    &10036  & 10069 &20020721  & -6.0  & 1429  &17.57\\
13    &10105  & 10069 &20020910  & -8.0  & 2028  &28.38\\
14    &10386  & 10365 &20030617  & -7.0  &  418  &54.31\\
15    &10464  & 10484 &20030926  &  4.0  &  817  &17.88\\
16    &10507  & 10488 &20031119  & 10.0  & 1190  &13.23\\
17    &10508  & 10486 &20031119  &-17.0  &  937  &25.08\\
18    &10520  & 10484 &20031216  &  1.0  &  232  &56.46\\
19    &10661  & 10652 &20040818  & 10.0  &  724  &24.04\\
20    &10708  & 10696 &20041203  &  8.0  &  219  &55.61\\
21    &10743  & 10735 &20050319  & -8.0  &  525  &40.70\\
22    &10750  & 10735 &20050408  & -7.0  &  205  &51.75\\
23    &10775  & 10792 &20050612  & 10.0  &  485  &31.27\\
24    &10803  & 10792 &20050825  & 12.0  &  272  &40.14\\
25    &10875  & 10865 &20060425  &-10.0  &  644  &32.90\\
26    &10904  & 10908 &20060812  &-13.0  & 1003  & 9.94\\
27    &10930  & 10923 &20061212  & -6.0  &  910  &29.36\\
28    &10935  & 10923 &20070104  & -6.0  &  365  &41.59\\
\hline\hline \label{tab:repeatingARs}
\end{tabular}
\end{center}
\end{table}

\end{document}